\documentclass[twocolumn,showpacs,preprintnumbers,amsmath,amssymb]{revtex4}
\usepackage[dvips]{graphicx}

\begin{document}
\title{Simulations of an ultracold, neutral plasma with equal masses}
\author{F. Robicheaux}
\affiliation{Department of Physics, Purdue University, West Lafayette,
Indiana 47907, USA}
\email{robichf@purdue.edu}
\author{B. J. Bender}
\author{M. A. Phillips}
\affiliation{Department of Physics, Auburn University, AL
36849-5311}
\date{\today}

\begin{abstract}
The results of a theoretical investigation of an ultracold, neutral
plasma composed of equal mass positive and negative charges are reported.
In our simulations, the plasma is created by the fast dissociation
of a neutral particle. The temperature of the plasma is controlled
by the relative energy of the dissociation. We studied the early
time evolution of this system where the initial energy was tuned
so that the plasma is formed in the strongly coupled regime. In
particular, we present results on the temperature evolution and
three body recombination. In the weakly coupled regime, we studied
how an expanding plasma thermalizes and how the scattering
between ions affects the expansion. Because the expansion causes
the density to drop, the velocity distribution only evolves for
a finite time with the final distribution depending on the
number of particles and initial temperature of the plasma.
\end{abstract}

\pacs{52.55.Dy, 34.80.Lx, 52.65.-y, 32.80.Ee}

\maketitle
\section{Introduction}

Ultracold and strongly coupled plasmas are non-traditional plasmas
where the strong interparticle interactions can lead to collective
behavior in the system.\cite{KPP} Depending on the system, highly correlated behavior can
emerge when the interaction energy exceeds the kinetic energy
because the particles can only move in the classically allowed region
of a highly structured potential energy surface. Experiments and
calculations of
ultracold plasmas consisting of electrons and positive atomic
or molecular ions have demonstrated a wide variety of interesting
effects.\cite{KPP}

In this paper, we present the results of calculations of ultracold
neutral plasmas where the masses of {\it all} particles are the
same. 
There have been many experiments and
simulations on electron-positron plasmas in many different situations (two
of many examples are Refs.~\cite{GS,RBD}). However, the ultracold
regime is largely unexplored. As discussed in Ref.~\cite{KRA},
a possible way to create an ultracold, equal mass plasma is to
dissociate cold molecules to a positive/negative ion pair (e.g., Rb$^+$
and Rb$^-$ with a ($M_--M_+)/(M_-+M_+)\sim 10^{-5}$).\cite{BBS}
The kinetic energy of the break up can be controlled by starting
in different vibrational states of the molecule. For example,
the Frank-Condon overlap region moves to larger $R$ as the vibrational quantum
number increases. With this picture in mind, we performed calculations
where many pairs of oppositely charged ions are launched with opposite
velocities so that the center of mass velocity of each pair is 0.
Each pair has a random center of mass position. When the positive and
negative ions separate to large distances, they can interact with all
of the other ions leading to a variety of effects.

One of the more interesting parameters that characterizes
the plasma behavior is the Coulomb coupling parameter
\begin{equation}
\Gamma =\frac{e^2/(4\pi\varepsilon_0 a_{ws})}{k_B T}
\end{equation}
where the Wigner-Seitz radius
\begin{equation}
a_{ws} = \left(\frac{3}{4\pi 2 n}\right)^{1/3}
\end{equation}
and $e$ is the electron charge, $k_B$ is Boltzmann's constant,
$n$ is the number density of one species,
and $T$ is the temperature; this expression has twice the density
of the usual formula because both species can interact with
the same footing. If $\Gamma$ is larger than $\sim 1$,
then the components of the plasma start displaying correlated
motion and that component of the plasma
is considered to be strongly coupled. If $\Gamma$
is less than $\sim 1$, then the plasma is weakly coupled. In
the weakly coupled limit, standard plasma concepts (e.g. Debye
screening, electron-ion scattering, ...) and standard atomic
concepts (e.g. three body recombination) are expected to be
good approximations.

We present results of two qualitatively different aspects of
an equal mass, ultracold neutral plasma. 
Previous calculations and experiments suggest that the early
time behavior\cite{KLK,KON,NGR} and the late time expansion of the plasmas
should be worth studying.\cite{KKB,MCK,RH1,KO1,KO2}

The early time behavior can be
interesting when the initial plasma temperature is low because
a strongly coupled plasma heats up due to the energy released
from three body recombination.\cite{KLK,MCK,RH1,KO1,KO2,KON,NGR}
For the typical ultracold plasma,
two electrons scatter in the vicinity of a positive ion leading
to one electron becoming bound while the second leaves with a
larger energy. As seen in Refs.~\cite{KON,NGR}, the electron
coupling parameter decreases from $\sim 1$ to $\sim 0.6$
on a time scale of $\sim 10$ plasma periods of the electron.
Reference~\cite{KON} presented calculations of
the early time heating of the plasma due to this process for
an ion mass substantially larger than the electron mass but much
less than physical masses. Reference~\cite{NGR} revisited this
system and calculated the heating as a function of the ion mass
(Ref.~\cite{NGR}, Fig.~2) and found that smaller ion masses gave lower electron
temperatures. These results motivated us to study the early time
behavior of initially strongly coupled plasmas when the mass of the
positive charge equals that of the negative charge. In addition to
the possible changes to the three body recombination and heating,
we point out that the ion plasma period would be of order
100$\times$ longer than the electron plasma period. Thus, there
would be an experimentally substantial time interval for studying neutral
plasmas where both components have $\Gamma\sim 1$.

The other possibility for interesting physics is in the expansion
of the plasma.\cite{KKB,RH1,KO1,KO2}
For this investigation, we used plasma temperatures sufficiently high
that we could treat the species as weakly coupled.
In the usual ultracold plasmas, the electrons are confined by the space
charge of the ions and the pressure from the electrons gives
a radially outward force on the ions.\cite{KLK,KPP} During the expansion,
the electron temperature drops substantially and this energy is
converted to radial kinetic
energy of the ions. For an equal mass plasma,
this description of the expansion can not be correct; unlike the
electron-ion plasma, the equal mass plasma could expand faster than
the plasma properties can be established. The plasma period
scales like $n^{-1/2}$ where $n$ is the density
whereas the expansion time scales with
$L/v_{th}\propto L T^{-1/2}$ where $L$ is a size scale of the plasma
and $T$ is the temperature of one of the components. Since these two
time scales vary with completely different parameters, it should be
possible to explore scenarios where there are many plasma oscillations
before the plasma substantially expands. Since the two species
start with the same temperature, the plasma expansion can not
extract energy from the thermal motion. The dominant effects arise
from the competition between particle scattering (which leads to
thermalization and slows the expansion)
and plasma expansion (which decreases the density and slows the
particle scattering).

In all discussions below, the quoted density is the density of one species.
The total density of particles is, of course, twice this value.

\section{Numerical Method}

We were interested in two qualitatively different types of behavior
which required two different types of computational methods.
For the early time behavior of a strongly coupled plasma, we
needed to simulate the thermalization and evolution of the
coupling parameter on a time scale less than 10 plasma periods.
This required a method that computes the positions and velocities
of all the particles. For the late time behavior, we needed to
simulate the thermalization and expansion of the plasma for the
case of weak coupling. This required a method that could handle
millions of particles over very long time scales but where
only pair-wise collisions are relevant.

\subsection{Molecular dynamics method}

The calculations for the early time behavior of a strongly
coupled plasma used an adaptive steps-size, Runge-Kutta
algorithm\cite{PTV} to solve the classical equations of motion
with wrap boundary condition for the forces. These calculations
used the same method and programs as described in Ref.~\cite{NGR}
Secs.~2 and 3 except the masses of the positive and negative
particles are the same. All of the particles
were contained within a cube of length $L$ so that the number of
positive particles divided by the volume equaled the target
density. When a particle reached the edge of the cube, it was
wrapped back into the cube; for example, if the $x_j$ was larger
than $L$ then it was replaced by $x_j-L$ before the next time
step. We did not use a pure Coulomb force for the particles all of
the way to zero separation. We derived the force from a spherically
symmetric potential between particles $i$ and $j$ that had the
form
\begin{equation}
PE_{ij}=\frac{q_iq_j}{4\pi\varepsilon_0\sqrt{r_{ij}^2+(Ca_{ws})^2}}
\end{equation}
where $q_i$ is the charge of particle $i$,
$r_{ij}$ is the separation
of particles $i$ and $j$, $a_{ws}=[3/(4\pi 2 n)]^{1/3}$ is the
Wigner-Seitz radius which is found from the number density $n$,
and $C$ is a constant substantially less than 1. We performed
calculations for several values of $C$ from $0.01$ to $0.05$ to
determine the effect that the soft core had on the dynamics. The
role that the constant $C$ plays in the calculation are discussed below
in the results of the different simulations. We emphasize
that this form of the potential is a numerical device for speeding
the calculation while keeping the physical result the same.
In our simulations, we are calculating the behavior for a finite
number of particles so the treatment of the boundary could be important.
As is typical, we used a wrap boundary condition on a cube
when computing the forces. When we computed the
force or potential energy between two particles, we would use
$x_i-x_j$ if $-L/2\leq (x_i-x_j)\leq L/2$ but would use
$x_i-x_j+L$ if it was less than $-L/2$ and $x_i-x_j-L$ if it
was greater than $L/2$. Similar definitions applied to the $y$-
and $z$-components. As the number of particles in the simulation
increases for a given density, the size of the cube increases and
the effect of the edges becomes less. We checked convergence
with respect to the cube size by comparing the results from different
size runs.

The particles were initialized to
mimic a sudden dissociation of positive and negative ions. Each
pair of positive and negative ions were randomly placed within the
cube. The initial separation of a positive-negative ion
pair was $a_{ws}/2500$; the separation vector for
each pair was randomly chosen with a uniform distribution on a sphere.
Their initial velocity vectors were chosen to be equal and opposite
so the center of mass velocity was zero; their directions were chosen
so they moved directly apart. The magnitude of velocity was chosen
so that the energy of the pair would be $2/3$ the temperature.

We made sure our reported results were converged with respect to
the time step in the calculation. This was checked by increasing the
accuracy parameter in the calculation. We checked that the results
in our plots did not change and we checked that the total energy
of the system drifted by less than a mK per particle.

\subsection{Fokker-Planck method}

We also wanted to investigate how an experimentally sized system
would thermalize and expand. For this situation, there could be
millions of particles but we studied cases where
the temperature was high enough that
the plasma was only weakly coupled. These conditions suggest
using a Fokker-Planck type method to include the effect of
scattering on the motion. We simulated a case where the positive
and negative ions have a Gaussian distribution in space but
each pair is launched back-to-back. The plasma will maintain
its spherical symmetry during the expansion.
Because there could be different conditions
in different radial regions we implemented a somewhat
complex Fokker-Planck method. The method was nearly
identical to that described for electron-electron scattering
in Sec.~IIIA of Ref.~\cite{RH2}.

As in the previous section, the positive and negative particles
are launched in pairs with 0 center-of-mass velocity. The pairs
are launched with a fixed separation and speed. In the
absence of scattering with other particles, {\it all} particles
would have the same speed. The initial speed distribution is
strongly peaked. Collisions broaden the distribution and, in the
limit that the particles can scatter many times, the velocities
will approach a Maxwell-Boltzmann distribution.
The ``temperature" quoted for each calculation is $2/3$ the
energy for each pair. The initial center of mass position for each pair is
chosen randomly from a spherical Gaussian distribution proportional
to $\exp (-r^2/L_g^2)$ where the length scale
\begin{equation}
L_g=(N/\bar{n})^{1/3}/\sqrt{2\pi }
\end{equation}
where $\bar{n}$ is the average density and $N$ is the number
of positive (or negative) ions. The local (single species) density is
$n(r)=2^{3/2}\bar{n}\exp (-r^2/L_g^2)$.

For the Fokker-Planck simulation, we would first step every
particle's position using it's velocity and then we would
update the velocity using pairwise scattering between particles.\cite{RH2}
In this method, the local density is used to determine the
scattering rate experienced by each particle
Unlike the electron-electron scattering in Ref.~\cite{RH2},
the ions in our plasma tend to stay near the same ions once
the expansion begins which led to numerical instabilities.
We avoided this problem by computing the plasma density
on a grid that evolved in time; the radial grid kept a
fixed number of points but the spacing changed with the
particle with the largest $r$: $\delta r=\max (r)/N_r$
where the number of grid points, $N_r$, was fixed.
As in Ref.~\cite{RH2}, only particles with nearly the same
$r$ are allowed to scatter from each other.

There are three important parameters to check for convergence.
The most important parameter to test is the time step. The
probability for a pair to scatter during a time step is
proportional to $\delta t$. If this probability becomes larger
than $\sim 0.1$ for a substantial fraction of the pairs, then
the effect from scattering will be underestimated. Another
parameter to check is the radial dependence of the density
which determines how often an ion near a radius $r$ will
scatter. We compute the density by distributing particles
on a radial grid. If the grid is too coarse, the density variation
will not be as rapid as it should. If the grid is too fine,
then the statistical noise from the finite number of particles
in the calculation will give errors in the local density.
Finally, we needed to only allow scattering between pairs
of particles with small separation in $r$. This was accomplished
in two steps. First a group of possible scatterers were randomly
picked from the particles. Second, a scattering particle was
picked and the particle in the scattering group with the
closest $r$ was used to scatter. Lastly, the particle pulled
from the scattering group was replaced by another random
particle. Ideally, the scattering group should consist of
all particles but it would be prohibitively slow to search
such a large list. We checked the convergence of the
results with the size of the scattering group by doubling
its size until the final results differed by less than
2\%.

To test whether this method could give reasonable results,
we performed molecular dynamics calculations with 6,400 positive
and 6,400 negative ions and compared to the Fokker-Planck
calculation. The comparison between these calculations are
given in the results section below.

\section{Scaling}

All of our calculations are purely classical which means that our
results have exact scaling properties. Therefore, we can present
our results for one specific choice of mass and density while
varying the initial energy. These results will be applicable to
other cases that have the same scaled parameters.

{\it Mass scaling.} All of the results presented below were obtained
for the positive and negative charges each having the mass equivalent
to one proton. To obtain the scaling relation, note that only the time
and speed
variables need to be scaled to obtain the same equations of motion.
If you write $t = \tilde{t}\sqrt{M/\tilde{M}}$,
you find that all of the velocities
are scaled by the factor $\sqrt{\tilde{M}/M}$ but {\it all} other parameters
(e.g. positions, energies, etc) are unchanged. As an example, if the
only change is that the mass is larger by a factor of 16 while the
initial density and energy are held fixed,
then the same motion occurs but over a time scale
a factor of 4 longer.

{\it Density scaling.} If the mass and charge are held fixed, the
classical equations of motion exactly scale under the transformation
\begin{equation}
v=\alpha^{-1}\tilde{v}\qquad r = \alpha^2\tilde{r}\qquad t=\alpha^3\tilde{t}
\end{equation}
where $\alpha$ is a dimensionless scale factor. The energy scales
as $E=\alpha^{-2}\tilde{E}$ while the density scales as $n=\alpha^{-6}\tilde{n}$.
Thus, we can explore different parameter regimes by either changing
the initial energy or the initial density; it is not necessary to change
both. The scaled energies and densities are related through
$n=\tilde{n}(E/\tilde{E})^3$ which means decreasing the energy
by a factor of 2 while keeping the initial density fixed is equivalent to 
increasing the initial density by a factor of 8 while keeping
the energy fixed.

\section{Results}

In this section, all of the calculations use a mass equal
to that for one proton and an average (one-species)
density of $n=10^9$~cm$^{-3}$. Since the masses are all equal,
we use $2n$ in all of the expressions for the plasma parameters.
The plasma frequency $\omega_p=\sqrt{2 ne^2/M\varepsilon_0} =
5.89\times 10^7$~rad/s and the Wigner-Seitz radius is
$a_{ws}=(3/4\pi 2n)^{1/3}=4.92$~$\mu$m. When computing the
Coulomb coupling constant
$\Gamma = e^2/(4\pi\varepsilon_0 a_{ws} k_B T)$ gives
$\Gamma = (3.39\; K)/T$. 

\subsection{Early time behavior of strongly coupled plasma}

In this section, we present the results of the evolution
of an ultracold plasma at early times. We performed
convergence checks as in Ref.~\cite{NGR}. We found our results
depended less on the soft core parameter $C$ than in
Ref.~\cite{NGR}; all of the results in this section
used $C=0.01$ although they were nearly indistinguishable
from the $C=0.02$ or 0.03.

For this section,
the particles are launched with an energy so that the average temperature
for a diffuse plasma would be 1~K.
The plasma should be strongly coupled with such a low temperature:
$\Gamma = 3.39$. However, ultracold
neutral plasmas consisting of electrons and ions do not exhibit
strong coupling of electrons due to heating from three body
recombination. References~\cite{KON,NGR} showed the temperature
rise at early times (of order 10 plasma periods); $\Gamma_e$
starts at $\sim 1$ due to disorder induced heating over a
time scale of $\sim 1/\omega_p$ followed by a decrease to
$\sim 0.6$ over a time scale of $\sim 10$ plasma periods. We
expect a similar effect for equal mass plasmas because three
body recombination should be important, but the role of the
mass {\it will} change the results.

\begin{figure}
\resizebox{80mm}{!}{\includegraphics{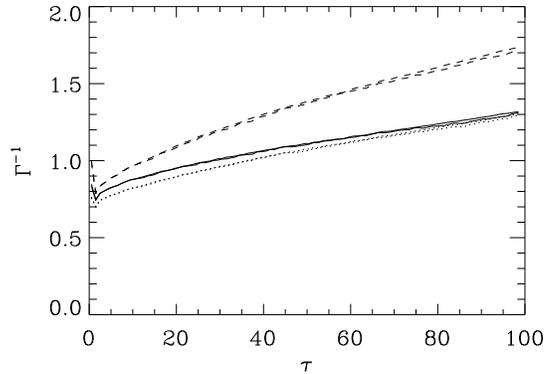}}
\caption{Six calculations of the scaled temperature,
$\Gamma^{-1} = k_BT/[e^2/(4\pi\varepsilon_0a_{ws})]$,
are shown as a function of the scaled time $\tau =\omega_p t$.
The plasma parameters are in the text. 
The solid lines are when an ion closer to another
ion than $0.1a_{ws}$ were excluded from the calculation of the temperature
and, for the dotted line, the
exclusion region was $0.2a_{ws}$. The dashed line includes all
ions, even those forming bound states.
Each line type is actually two lines: one for a calculation using
800 particles of each sign and one for a calculation using 400
particles of each sign. The near equality of the calculations with
400 and 800 particles demonstrates the convergence.
}
\end{figure}

Figure~1 shows the
inverse of the coupling constant with time scaled by the plasma frequency.
The $\Gamma^{-1}$ is proportional to the temperature while
quickly showing whether the plasma is strongly coupled. We defined
the temperature to be 2/3 the average kinetic energy. One difficulty
is in deciding which particles to include in the average kinetic
energy. Reference~\cite{KON} took all
electrons that were not deeply bound to an ion and found their
velocity distribution; they fitted this distribution to a
Maxwell-Boltzmann distribution to obtain the temperature. Reference~\cite{NGR}
obtained the temperature by using the equipartition theorem
for electrons that were further than a specified distance from
every ion; the distance we chose was either $a_{ws}/10$ or
$a_{ws}/5$ (comparing the two calculations gives an estimate
of the uncertainty of the temperature);
they found this result agreed with the method of Ref.~\cite{KON}.
In Fig.~1, the dotted and solid lines correspond to the scaled
temperature when the distance was chosen to be $a_{ws}/5$ and
$a_{ws}/10$ respectively. The dashed line includes all particles
but gives an unphysically high temperature due to ``deeply" bound
ion pairs.

The value of the plasma period can be used to convert the time
axis to physical value. For protons, the final time is $\simeq 1.7$~$\mu$s.
For other ions, the time scale would increase by a factor of
$\sqrt{M_{ion}/M_{proton}}$ which could easily stretch the time
scale to order 10$~\mu$s. This would be easily within the time
scale of experimental probes (e.g.~see Ref.~\cite{CSL}).

A Coulomb coupling parameter of $\Gamma = 1$ is approximately the
demarcation between a strongly or a weakly coupled plasma. The
results in Fig.~1 are similar to those for electron-ion plasmas
(e.g.~Fig.~1 of Ref.~\cite{NGR}) but the
coupling parameter for equal mass particles
is somewhat larger than the electron-ion case. However,
the electron component of the ultracold plasma
is hard to probe and the total time scale for a duration like Fig.~1
would be $\sim 60$~ns. Although the results for the equal mass
plasma do not qualitatively differ from the electron-ion case,
the equal mass plasma is worth studying because
it seems possible to experimentally probe the
equal mass plasma with parameters near that for strong coupling.

\begin{figure}
\resizebox{80mm}{!}{\includegraphics{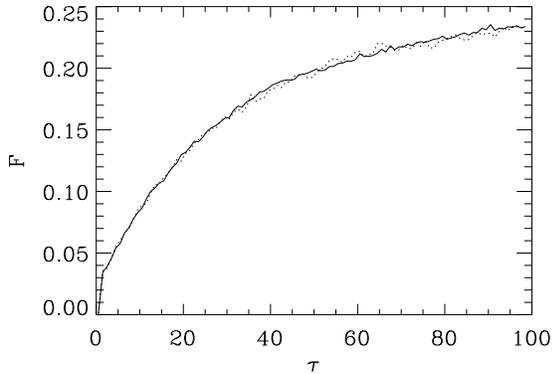}}
\caption{Fraction of bound particles
as a function of the scaled time $\tau =\omega_p t$.
The solid line used 800 particles of each type while the
dotted line used 400 particles of each type.
}
\end{figure}

The heating in Fig.~1 is due to three body recombination, but the
number of bound atoms can not be inferred from this data.
Figure 2 shows the fraction of bound particles as a function of
time for the plasma parameters of Fig.~1.
For this plot, we defined a pair $j$-$j'$ as bound if the
closest particle to $j$ was $j'$ for two times separated by
$1/\omega_p$ {\it and} the separation was less than
$a_{ws}/5$. The results did not substantially
change when the separation distance was less than
$a_{ws}/10$.

Figure~2 shows that a substantial fraction of particles become
bound over this time range. As might be expected, the fraction
of bound pairs has a rapid initial increase followed by a much
slower rise. This is expected because the three body recombination
rate rapidly decreases with increasing temperature and is proportional
to the square of the number of free particles. If the density drops
by a factor of 0.8, then the recombination rate drops by the factor
0.64.

The center-of-mass speed of the bound pairs will be substantially
less than that of the free particles. Therefore, it should be possible
to detect the bound pairs by waiting for the free particles to
expand out of the region where the plasma was created. Later, the
particles can be dissociated and detected by ramping on an electric
field.

In Fig.~1, we did {\it not} include the change to the Wigner-Seitz
radius, $a_{ws}$, in the changing $\Gamma$. However, this change
does not have a large effect. Changing the density by a factor of
0.77 changes $a_{ws}$ by a factor of 1.09. This would increase the
$\Gamma^{-1}$ by the same factor of 1.09.

\subsection{Expanding Plasma}

\begin{figure}
\resizebox{80mm}{!}{\includegraphics{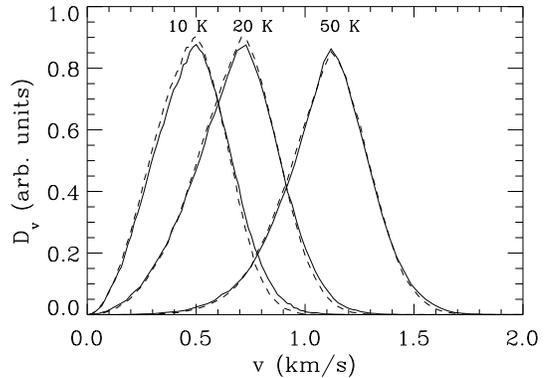}}
\caption{The velocity distribution for an expanding plasma
composed of 6,400 positive and 6,400 negative ions for
three different temperatures. The final time for all
calculations was 4 plasma periods, i.e. $t_{fin}=8\pi /\omega_p$.
The solid line is the result from the molecular dynamics
calculation and the dashed line is the result of the
Fokker-Planck calculation. The individual temperatures have
been scaled to a peak of $\sim 0.9$ but the relative scaling
of the Fokker-Planck and molecular dynamics is held fixed.
}
\end{figure}

This section contains our results on how the ion-ion scattering
affects the plasma expansion. If there were no scattering, then
every ion would have the same speed. Since the particles start in
a compact region, they would move approximately radially at late times
with a speed $v_{init}$. This would give a radial shell of particles
expanding with this speed. Collisions between ions could drastically
change this picture. There is an interesting competition between
scattering and plasma expansion. The scattering tends to delay
the expansion. However, once the expansion is underway, the scattering
rate quickly drops. Therefore, there is a well defined final velocity
distribution that depends on the initial number and temperature
of the particles. Since Coulomb collision rates decrease with temperature,
the effects from scattering will be most pronounced at the lowest
temperatures.

The collisions will tend to randomize the
velocities, with more collisions tending to give a larger spread
of velocities. In the limit of infinite collisions, the velocity
distribution will go to a Maxwell-Boltzmann distribution. In terms
of the distribution of speeds, the Maxwell-Boltzmann distribution
is proportional to $v^2\exp (-Mv^2/[2 k_BT])$ which has a peak at
$v = \sqrt{2 k_BT/M}$. The initial peak in speed is at
$v_{init}=\sqrt{3 k_BT/M}$. Thus, the collisions will move the peak
in the distribution to smaller speeds while broadening the distribution.
Another effect is that each collision changes the direction the particle
is traveling. This will slow the initial expansion of the plasma because
the particle motion will be more like a random walk. This slowing effect could
be observable if the particle detector is close to the plasma because
the delay could be a large fraction of the travel time.

Performing molecular dynamics calculations for $10^5-10^6$ particles
would be incredibly slow which is why we used a Fokker-Planck method
to obtain the results in this section. Figure~3 shows a comparison
between the two methods for a small enough number of particles where
both methods can be used. The initial speed for 10, 20, and 50~K
are 498, 704, and 1114 m/s respectively. All three distributions
are peaked near these values but are substantially broadened. If these
distributions were Maxwell-Boltzmann, the peaks would be at
407, 575, and 909~m/s respectively. As expected the 10~K distribution
shows the largest spreading relative to the peak.

Most important is the comparison between the two methods. There is
good agreement for the three different temperatures. The best
agreement is for 50~K which we expected since this plasma is the best
example of a weakly coupled plasma. We did not test calculations for
temperatures lower than 10~K because the Coulomb coupling is becoming
uncomfortably large, $\Gamma = 0.34$, at this temperature. In fact,
the 10~K results have the largest disagreement. The Fokker-Planck
calculations somewhat underestimate the high energy part of the
distribution. However, this difference seems to be small enough that
we can use this approximation to obtain the general behavior of the
plasma expansion.

\begin{figure}
\resizebox{80mm}{!}{\includegraphics{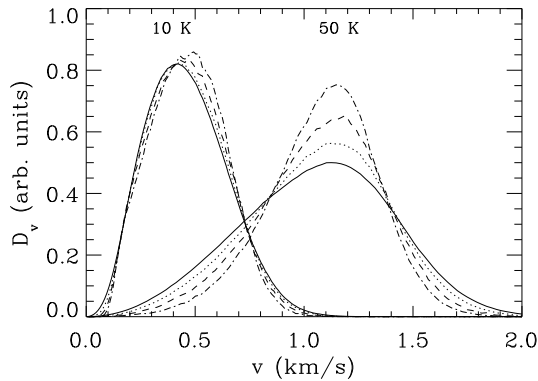}}
\caption{The velocity distribution for an expanding plasma
for different number of ions for
two different temperatures. The final time for all
calculations was $6 L_{g,max}/v_{init}$ where $L_{g,max}$
was the $L_g$ for the calculation with $5.12\times 10^6$
particles of each type.
The solid line is for $5.12\times 10^6$, the dotted line is for
$1.28\times 10^6$, the dashed line is for $3.2\times 10^5$
and the dot-dash line is for $8\times 10^4$ particles of
each type. The individual temperatures have
been scaled but the relative scaling
within a temperature is held fixed.
}
\end{figure}

Figure~4 shows the velocity distribution for different numbers
of particles. The final time in all of the calculations is large
enough that the plasma has expanded to the point where
further changes in the velocity distribution are negligible.
Each line type corresponds to a factor of 4 in the number of
particles. The largest calculation had 5.12 million particles of
each type while the smallest had 80 thousand. Note that even
the smallest calculation has 10 times more particles than in
Fig.~3. There are a few general trends worth noting.

One of the interesting features is that the peak of the 10~K
distribution is moving to smaller $v$ while the position of the peak of the
50~K distribution is nearly unchanged. In Fig.~3, the peak for
all of the distributions was near the $v_{init}$ for that temperature.
It appears that the scattering leaves the position of the peak approximately
unchanged until the width of the distribution becomes nearly
equal to the value of the peak. Another interesting feature is
how little the distributions change as more particles are added.
Each line type corresponds to a factor of 4 in the number of particles
which is a factor of $4^{1/3}\simeq 1.6$ in the size of the plasma.
If the plasma size increases by a factor of 1.6, the amount of time
when collisions could happen increases by at least a factor of 1.6.
(Since collisions slow the plasma expansion, the duration for possible
collisions will increase somewhat faster than linearly with the
plasma size.) Although each line type corresponds to another factor
of 1.6 in plasma size, there is only $\sim 10$\%
change in the 50~K distribution; the almost complete lack of change
in the 10~K distribution, for the largest numbers, will be addressed
below.

The speed distribution must have small $v$ behavior proportional to
$v^2$ or a higher power. This masks one of the trends that arises
because Coulomb scattering is larger at smaller relative energy.
The high energy tail of the distribution is being filled more
slowly than the low energy part. This is easiest to see in the
5.12~million calculation for 50~K (solid line). The distribution
drops from the peak faster on the high energy side than the low
energy side.

\begin{figure}
\resizebox{80mm}{!}{\includegraphics{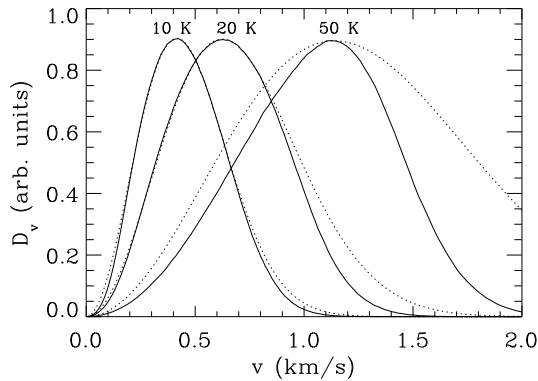}}
\caption{The velocity distribution for an expanding plasma
for three different temperatures. The final time was the same
as Fig.~4. The Fokker-Planck calculations are the solid lines
and have $5.12\times 10^6$ particles of
each type. The dotted lines are Maxwell-Boltzmann distributions
chosen to have a peak at the $v$ from the Fokker-Planck calculation.
All curves have been multiplied by a scaling constant so they
all have the same height.
}
\end{figure}

Figure~5 shows a comparison of the final speed distribution
to a Maxwell-Boltzmann distribution. All of the Fokker-Planck
calculations have $5.12\times 10^6$ particles of
each type and, thus, should have the largest amount of
scattering of our calculations. As expected, the 10~K distribution
is most similar to the Maxwell-Boltzmann distribution while the
50~K distribution is least similar. This shows that the distribution
does approach thermal for a large enough number of particles.
The small changes in the 10~K distribution in Fig.~4 is because
the distribution is nearly thermal.

The 20~K and 50~K distributions both show that the difference
from a thermal distribution is more strongly pronounced on the
high energy side of the peak compared to the low energy side
because Coulomb collisions thermalize high energy particles
slower than low energy particles.

Experimental results corresponding to Figs.~4 and 5 would be
interesting. These plots are, in essence, snapshots of how
the Maxwell-Boltzmann distribution is reached in a plasma.
By changing the number and/or energy of the plasma, there is
direct control over the duration of the particle scattering. The
final velocity distribution gives a direct test of our understanding
of charged-particle scattering in a plasma.

\section{Conclusions}

We have performed two types of calculations for ultracold neutral plasmas
where all of the particles have the same mass. In our molecular
dynamics calculations, we investigated the initial heating and
formation of bound states for a plasma that is strongly coupled
initially. We found similar heating as observed in ion-electron
calculations but the physical time scale is now of order 1-10~$\mu$s.
In our Fokker-Planck calculations, we studied the expansion and
thermalization of particles of a weakly coupled plasma. Because
the density drops with expansion, the collision rate rapidly drops
and the velocity distribution stops evolving. We showed that the
speed distribution approaches a Maxwell-Boltzmann distribution
at low temperature and with enough particles; the largest differences
are on the high energy side of the distribution.

There are several situations that we did not investigate due to
the limitations of our computational tools. One interesting possibility
would be to study this system in a strong magnetic field.\cite{ZFR} If the
cyclotron radius of the motion becomes comparable to or smaller than the
plasma size, there should be interesting modifications to the scattering.
Also, there could be interesting plasma waves and instabilities because
both species have the same mass which would argue that a plasma wave
in one species will be degenerate with one in the other species.
Another interesting possibility would be to study the expansion of
a plasma that is initially strongly coupled. A final interesting
possibility would be to include a direction distribution for the
initial dissociation of charges; for example, there would be
different scattering if the charges are launched mainly in the
$\pm z$-direction than if they are launched in random directions
as was done in our calculations.

We thank Phil Gould for interesting discussions that motivated
us to study this system.
This material is based upon work supported by the U.S. Department
of Energy Office of Science, Office of Basic Energy Sciences
Chemical Sciences, Geosciences, and Biosciences Division
under Award Number DE-SC0012193.

\end{document}